\documentclass[prl,superscriptaddress,showpacs,twocolumn]{revtex4}
\usepackage{amsmath,graphicx}
\DeclareMathOperator{\re}{Re}
\DeclareMathOperator{\im}{Im}
\begin{document}
\title {Yang-Lee zeroes for an urn model for the separation of sand}
\author{I. Bena}
\affiliation{Department of Physics, University of Geneva, CH 1211
Geneva 4, Switzerland}
\author{F. Coppex}
\affiliation{Department of Physics, University of Geneva, CH 1211
Geneva 4, Switzerland}
\author{M. Droz}
\affiliation{Department of Physics, University of Geneva, CH 1211
Geneva 4, Switzerland}
\author{A. Lipowski}
\affiliation{Department of Physics, University of Geneva, CH 1211
Geneva 4, Switzerland}
\affiliation{Faculty of Physics, A.~Mickiewicz University,
61-614 Pozna\'{n}, Poland}

\begin {abstract}
We apply the Yang-Lee theory of phase transitions to an urn model of separation of sand. The effective partition function of this nonequilibrium system can be expressed as a polynomial of the size-dependent effective fugacity $z$. Numerical calculations show that in the thermodynamic limit, the zeros of the effective partition function are located on the unit circle in the complex $z$-plane. In the complex plane of the actual control parameter certain roots converge to the transition point of the model. Thus the Yang-Lee theory can be applied to a wider class of nonequilibrium systems than those considered previously.
\end{abstract}
\maketitle

\textit{I. Introduction.}--- Nonequilibrium phase transitions have recently attracted increasing 
theoretical interests. One of the motivations is the idea that some of the concepts of equilibrium phase transitions might apply to nonequilibrium situations as well. In particular, there have been recent investigations~\cite{revBl} in the applicability of the Yang-Lee theory to nonequilibrium systems. Usually, those systems are modeled in terms of a master equation. Most of the physics is contained in the prescribed transition rates. When varying the control parameters, the system may exhibit a nonequilibrium phase transition in its stationary state. There are very few models for which such transitions can be described analytically. Moreover, it is not obvious to find a nonequilibrium function playing a similar role to the equilibrium partition function and for which the Yang-Lee strategy might be extended. Two families of dynamic models have been discussed so far in the
literature~\cite{revBl}.

(i) Most of the works concern driven diffusive systems~\cite{arndt,blythe,jafarpour}. These models have a
unique steady state, for which one can compute exactly the probability distribution of microstates. The normalization factor, defined as the sum over these probabilities, is shown to play the same role as the partition function in equilibrium systems. It is a polynomial in the control parameter, the latter being a constant transition rate, i.e., a size-independent quantity. The roots of this polynomial in the complex plane of the control parameter accumulate, in the thermodynamic limit, at the transition point. As in the case of equilibrium transitions, the way these zeroes accumulate nearby the real axis informs about the type of transition (first order if the line of zeroes is perpendicular to the real axis, and second order if it forms an angle of $\pi/4$ with this axis).

(ii) Other studies have been performed for directed percolation models on a square lattice in $(1+1)$ 
dimensions~\cite{dammer}. It was suggested that survival probability $P(t)$ for a finite propagation time $t$ might 
play the role of a partition function in this case, with the occupation probability $p$ as the complex control parameter. However, the survival probability, that is actually the order parameter in this problem, does not have the ``standard" properties of an equilibrium partition function~\cite{revBl}. Therefore it is not surprising that the distribution of its complex zeroes has a more complicated structure. It was also shown that the minimal distance $d(t)$ between the complex zeroes of the survival probability and the critical point $p_c$  scales for large $t$ as $t^{-1/\nu_{\parallel}}$, where $\nu_{\parallel}$ is the critical exponent for the temporal correlation
length.

The above examples suggest that it is rather straightforward to  apply 
Yang-Lee theory to nonequilibrium phase transitions, at least as long 
as one can construct the appropriate effective partition function (EPF). However, an implicit assumption of the above approaches is that the EPF can be expressed as a polynomial of a size-independent control parameter. 
The purpose of our letter is to show, on a simple model, that this should not be necessarily the case. Indeed, our model has an EPF that has a polynomial structure in terms of a \textit{size-dependent} effective fugacity, and not in terms of some size-independent parameter. The applicability of the Yang-Lee approach to such a case is thus highly questionable. We show that the Yang-Lee strategy for this model still works and thus can be generalized to a wider class of nonequilibrium phase transitions.

This Letter is organized as follows. In Sect.~II the model is defined and the EPF is introduced as a polynomial of an effective size-dependent fugacity. In Sect.~III, the zeroes of the EPF are numerically studied. They are shown to form a much more complicated structure in the plane of the complex fugacity than in the above discussed examples (i) and (ii). Nevertheless, it is still possible to extract some information about the nature of the phase transition in our model by investigating the zeroes in the complex plane of the control parameter. In particular, a subset of zeroes converges to the transition point. Moreover, the convergence seems to obey standard finite-size dependence arguments. In the Conclusions (Sect.~IV), we argue that the behavior of zeroes in our model is an indication of difficulties that might be typical to more general nonequilibrium systems, in particular those where transition rates between microscopic configurations are not constants (i.e., size-independent quantities), but state dependent.

\textit{II. The model.}--- The model we consider was introduced to describe spatial separation of vibrated sand~\cite{adam} and is a generalisation of Ehrenfest's urn model~\cite{ehrenfest}. 
In this model $N$ particles are distributed between two urns, the first urn containing $M$ particles and the second one $N-M$. The dynamics is defined as follows. At each time step, one particle is chosen at random in one of the urns.
Then, with a probability that depends on the number of particles present in the chosen urn, i.e., with a \textit{state-dependent} transition rate, this particle moves to the other urn. Correspondingly, the flux $F(n)$ of particles leaving a given urn at a certain time depends on the fraction $n$ of the total number of particles in the given urn at that moment. This model is thus by construction mean-field like. The master equation for the probability distribution $p(M,t)$ that there are $M$ particles in a given urn at time $t$ writes~\cite{adam}:
\begin{multline}
p(M,t+1)=F \left( \frac{N-M+1}{N} \right)p(M-1,t) \\
+ F \left( \frac{M+1}{N} \right) p(M+1,t) \\
+ \left[ 1-F\left(\frac{M}{N} \right) - F \left( \frac{N-M}{N} \right) \right]p(M,t).
\end{multline}
Its stationary solution is found to be~\cite{shim}:
\begin{equation}
p_{s}(M)= \frac{1}{Z_N} \prod_{i=1}^N \frac{F\left( \frac{N-i+1}{N} \right)}{F\left( \frac{i}{N} \right)},
\end{equation}
where the normalization factor is:
\begin{equation}
Z_N=1+
\sum_{M=1}^N\prod_{i=1}^N \frac{F\left( \frac{N-i+1}{N} 
\right)}{F\left( \frac{i}{N} \right)}.
\label{zn1}
\end{equation} 
In the spirit of models of type (i), we shall assume that $Z_N$ plays the role of an EPF.

This model describes the transition between a symmetric distribution of the particles in the two urns, associated with a single peak of the probability distribution at $M=N/2$ (for $N$ even), to a symmetry breaking state described by a bimodal distribution with peaks at $M=N(1/2 \pm \varepsilon)$. The order parameter $\varepsilon$ measures the difference in the occupancy of the two urns. To produce this symmetry breaking it suffices that the flux function $F(n)$ has a single hump~\cite{eggers,adam}. The simplest possible choice for $F(n)$ having this property is
\begin{equation}
F(n)=n \exp \left(-A \, n \right),
\end{equation}
which corresponds to a \textit{state-dependent} transition rate $\exp(-A n)$. Thus the problem is characterized by a single control parameter $A$. In the thermodynamic limit $N \to \infty$, this symmetry breaking corresponds to a second-order nonequilibrium phase transition. In this limit, the probability
distribution becomes $\delta$-peaked around the macroscopic stable state, that is determined by the condition that the flux of particles directed from the first urn to the second one equals the flux of particles 
from the second urn towards the first one, $F\left( 1/2 -\varepsilon \right) = F\left( 1/2 +\varepsilon\right)$. It follows that in the thermodynamic limit for $A<A_c=2$ the stationary
state is the symmetric one, while for $A>A_c=2$ the equipartition of particles is broken, i.e., a second 
order nonequilibrium phase transition takes place at $A=A_c=2$. 

\textit{III. Analysis of the zeroes of the EPF.}--- With such a choice of the flux $F(n)$, one may rewrite 
the normalization factor~(\ref{zn1}) as:
\begin{equation} 
Z_N=\sum_{M=0}^N \binom{N}{M} z^{M(N-M)}.
\label{zn2}
\end{equation}
Here $\binom{N}{M}=N!/[M!(N-M)!]$ is the binomial coefficient and $z=\exp\left(- A/N \right)$ is the effective fugacity. One can see that $Z_N$ is a polynomial in $z$, that is related to the control parameter $A$ of the model, but $z$ \textit{is not a size-independent quantity}, and depends on the number of particles $N$.

\begin{figure}
\begin{center}
\includegraphics[width=\columnwidth]{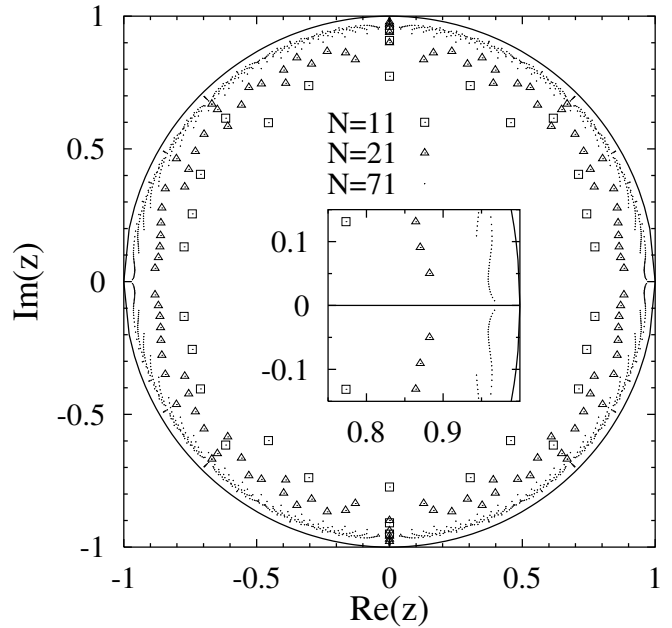}
\end{center}
\caption{Zeroes of the EPF~(\ref{zn2}) in the complex $z$-plane for three values of 
$N$. The continuous line is the unit circle. The inset illustrates the
behavior in the vicinity of $z=1$.}
\label{fig1}
\end{figure}

Nevertheless we embark on studying zeroes of the EPF $Z_N$. As a first step we find zeroes of Eq.~(\ref{zn2}), considering $z$ as a complex $N$-independent variable. The results of our numerical calculations, using \textsc{Mathematica}, for three values of $N$ are represented in Fig.~\ref{fig1}. Note that the order of the polynomial of Eq.~(\ref{zn2}) increases rapidly like $N^2/4$, and therefore we were not able to perform precise calculations of the roots beyond $N=71$. 

With increasing $N$ these roots approach the unit circle. One can argue that this should indeed be the case. First, let us associate with the EPF (\ref{zn2}) the nonequilibrium complex free energy density $f_N(z)= (1/N) \ln (Z_N)$. For large $N$ and $|z| > 1$ the EPF is dominated by the central term $M=N/2$ and hence $f_N(|z|>1) \sim (N/4) \ln z$. Thus we define $f^{(1)}(z) = (N/4) \ln z$ as a free energy in the $|z|> 1$ region. On the other hand, for $|z|<1$ the dominant contribution are coming only from the $M=0$ and $M=N$ terms, and thus $f_N(|z|<1) \sim 1/N$ which in the thermodynamic limit defines $f^{(2)}(z) = 0$ for $|z| < 1$. The above analysis indicates that the model for which Eq.~(\ref{zn2}) is the EPF with $z$ as a control parameter undergoes an abrupt transition at $|z|=1$. In the thermodynamic limit $N \to \infty$, $f_N$ exhibits an infinite jump at the transition. To obtain the location of the zeros of $Z_N$ in this limit we have to equate real parts of the nonequilibrium complex free energies on both sides of the transition~\cite{grossmann}. Namely, we require that
\begin{equation}
\re f^{(1)}(z) = \re f^{(2)}(z). \label{realf}
\end{equation}
Using a polar representation $z=r{\rm e}^{i\phi}$ we immediately obtain that the only way to satisfy Eq.~(\ref{realf}) is to have $r=1$. Hence, asymptotically, the zeroes should be located on the unit circle, as confirmed by our numerical calculations.

\begin{figure}
\begin{center}
\includegraphics[width=\columnwidth]{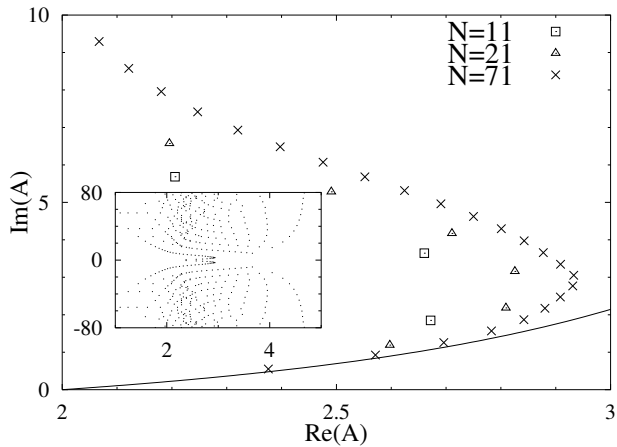}
\end{center}
\caption{Zeroes of Eq.~(\ref{zn2}) in the complex $A$-plane, nearby the 
critical value $A_c=2$, for three values of $N$. 
The inset shows more roots for $N=71$. The continuous line is the analytical 
perturbative estimation~(\ref{eq18}) of the line of zeroes in the thermodynamic limit, see
main text.}
\label{fig2}
\end{figure}

However, the model with $z$ as a control parameter (which has a transition with a jump of the effective free energy density) is quite different from the original urn model with $A$ as a control parameter (which has a continuous phase transition). Therefore, in order to infer some information about the phase transition in the urn model we have to analyze the behavior of zeros of Eq.~(\ref{zn2}) in the complex $A$-plane, that can be obtained from the zeroes in the $z$-plane using the relation $A=-N \ln (z)$. Transformation of zeros into the complex $A$-plane is shown in Fig.~\ref{fig2}.

With increasing $N$ the zeroes approach the critical point $A_c = 2$ with a slope that is close to $\pi/4$, and with a vanishing density of zeroes. These numerical observations seem to confirm 
the second-order nature of the phase transition~\cite{revBl}.
 
In the following we establish analytically the form of
the line of zeroes close to the critical point,  
and the final result is presented as a continuous line 
in Fig.~\ref{fig2}. As already mentioned, in the thermodynamic limit the EPF 
is dominated by the stationary state:
\begin{multline}
\left(1/2-\varepsilon\right) \exp \left[-A\left(1/2-\varepsilon\right)\right] \\
= \left(1/2+\varepsilon\right) \exp 
\left[-A\left(1/2+\varepsilon\right)\right]. \label{stat1}
\end{multline}
Below the critical point the leading term 
of $Z_N$ for $N \to \infty$ is given by the central peak
$M=N/2$:
\begin{equation}
Z_{N} \sim \binom{N}{N/2} \exp \left(-A \frac{N}{4} \right), \qquad A<A_c=2.
\end{equation}
On the other hand, for $A > A_c$ there are two leading contributions to $Z_N$
coming, respectively, from   
$M=N(1/2-\varepsilon)$ and $M=N(1/2+\varepsilon)$, where $\varepsilon$ is the
solution of the macroscopic stationarity condition (\ref{stat1}). Therefore:
\begin{multline}
Z_{N} \sim 2 \binom{N}{N(1/2-\varepsilon)}
 \exp \left[-A N \left( \frac{1}{4} - \varepsilon ^2 \right) \right], \\
A>A_c=2.
\end{multline}

Correspondingly, the effective free energy density $f=\lim_{N \to \infty}(1/N) \ln (Z_N)$~\cite{grossmann} associated with this EPF is
\begin{subequations}
\begin{eqnarray}
f^{(1)} &=& -A/4+ \ln 2, \; A<A_c=2, \\
f^{(2)} &=& -A(1/4-\varepsilon^2) - (1/2-\varepsilon) \ln (1/2-\varepsilon) \nonumber \\
& & -(1/2+\varepsilon) \ln (1/2+\varepsilon), \; A>A_c=2.
\end{eqnarray}
\end{subequations}

Let us now consider the behavior of the EPF and of the effective free energy density as a function of 
the \textit{complex} parameter $A$. Then the condition~(\ref{realf}), $\re f^{(1)} = \re f^{(2)}$, together with Eq.~(\ref{stat1}) determine the line of zeroes in the complex $A$-plane. Note that now $\varepsilon$ is a complex variable obtained from the steady-state Eq.~(\ref{stat1}).

However, Eq.~(\ref{realf}) is now too complicated to allow for a complete analysis of the  zeroes line in the entire $A$-plane. But we are mainly interested in the behavior of this line in the vicinity of the critical point $A_c=2$. Therefore, we shall look for a perturbative solution of Eqs.~(\ref{realf}) and (\ref{stat1}) in the small real parameter $\alpha =\re A - 2$ around $A_c$. An inspection of these equations
shows that $\im A$ scales like $\alpha$, while both $\re \varepsilon$ 
and $\im\varepsilon$ scale like $\alpha^{1/2}$. We are thus led to consider
the following developments:
\begin{subequations}
\begin{eqnarray}
A &=& (2 + \alpha) + i \alpha (a_0+\alpha a_1+\alpha^2 a_2+ \ldots), \label{series1} \\
\varepsilon &=& \alpha^{1/2} (x_0+\alpha x_1+\alpha^2 x_2+ \ldots) \nonumber \\
& & + i \alpha^{1/2} (y_0+\alpha y_1+\alpha^2 y_2+ \ldots).
\end{eqnarray}
\end{subequations}

We substitute these expressions in Eqs. (\ref{realf}) and (\ref{stat1}), then solve them order by order in $\alpha$. This leads to the following parametric expression for the line of zeroes in the $A$-plane, in the vicinity of $A_c=2$:
\begin{equation}
A=(2+\alpha) + i \alpha \left[1 + 0.6 \alpha + \mathcal{O}(\alpha^2)\right]. \label{eq18}
\end{equation}
The result of this perturbative calculation up to $\mathcal{O}(\alpha^5)$
is represented 
by the continuous line in Fig.~\ref{fig2}. Note that, indeed, the slope of this
curve at the critical point is $\pi/4$, sign of a second-order phase transition.
Moreover, one can compute the density of zeroes on this curve 
in the vicinity of the critical point using the relationship~\cite{grossmann}:
\begin{equation}
2\pi\mu(s)= \left| \frac{\partial}{\partial s}\im \left( f^{(1)}-f^{(2)} \right) \right|,
\end{equation}
where $\mu(s)$ is the density of zeroes at a distance $s$ from the transition point, distance measured along the line of zeroes. This gives in the vicinity of the transition point $\mu(s) = a s + \mathcal{O}(s^2)$, with $s=\sqrt{2}\alpha$ and $a=0.0045(1)$. The density of zeroes vanishes as a power law towards the transition point on the real axis -- i.e., we recovered yet another characteristic of the equilibrium theory for second-order phase transitions.

\begin{figure}
\begin{center}
\includegraphics[width=\columnwidth]{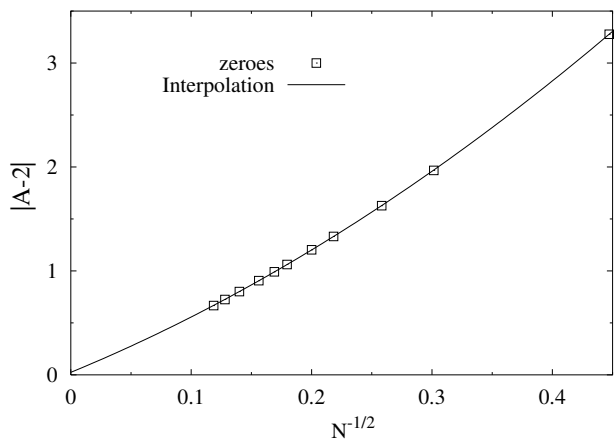}
\end{center}
\caption{The minimum distance between the zeroes of $Z_N$ and $A_c=2$ in the complex $A$-plane as a function of ${N}^{-1/2}$. The continuous line is a least-square fit of the form $a+b N^{-1/2}(1+c N^{-1})$, where $a,\ b$, and $c$ are fitting parameters. Note that the correction term $N^{-3/2}$ is used in view of the small values of $N$ that are accessible to calculations. Extrapolating to the $N\to \infty$ limit we obtain $a/A_c \simeq 1\%$, i.e., very close to zero, that confirms the theoretical value $A_c=2$.}
\label{fig3}
\end{figure}

Up to now, we have shown that the zeroes of the EPF indeed provide information about the location and the type 
of the phase transition. But, as pointed out~\cite{dammer} certain critical exponents are also encoded 
in the behavior of these zeroes. A similar conclusion can be drawn in our model. As shown in Fig.~\ref{fig3}, the distance $|A-A_c|$ between the closest root $A$ to $A_c=2$ in the complex $A$-plane decreases like $N^{-1/2}$. A simple scaling argument shows~\cite{dammer} that $|A-A_c|$ should scale with the system size as $N^{-1/\nu}$, where $\nu$ is the correlation length critical exponent. However, our urn model is structureless and the correlation length does not seem to be a well-defined quantity. Nevertheless, we can implement a definition of the critical exponent $\nu$ that is based on the finite-size scaling of moments of the order parameter~\cite{binder}. Since the probability distribution at the critical point for urn models is known~\cite{shim}, using the standard prescription~\cite{binder} we obtain $\nu=2$. Such a value agrees with the finite-size scaling observed in Fig.~\ref{fig3}.

\textit{IV. Conclusions.}--- We considered  a simple stochastic model with \textit{state-dependent transition rates} for which the EPF can be computed analytically and that exhibits a nonequilibrium second order phase transition. Our aim was to show that, although it is not  a straightforward task to apply the concepts of the Yang-Lee theory, it is a remarkable and non-trivial fact that they still apply when the microscopic transition rates of the model are state-dependent. Indeed, all  the other models that have been studied so far allowed for a rather simple transposition of the equilibrium theory. This was essentially due to the existence of some state-independent transition rates, and the possibility to express the EPF as a polynomial of these parameters. Our work therefore opens new perspectives for the study of other nonequilibrium models with phase transitions, indicating that Yang-Lee's ideas apply to a much wider range of phenomena than those considered up to now.
 

\end {document}